\documentclass{PoS}

\title{Latest CMS results on Higgs boson production in association with top quarks ($\mathrm{t\bar{t}H}$)}

\ShortTitle{Latest CMS results on  $t\bar{t}H$}

\author{\speaker{Saranya Samik Ghosh on behalf of the CMS Collaboration}\\
        CEA Paris-Saclay\\
        E-mail: \email{saranya.ghosh@cern.ch}}

\abstract{The searches for Higgs boson production in association with a pair of top quarks at the Compact Muon Solenoid (CMS) experiment using data collected during proton-proton collisions at a center of mass energy of 13 TeV in 2016 are presented. The decay modes of the Higgs boson that are discussed here correspond to the $H \rightarrow ZZ^{*}$, $H \rightarrow WW^{*}$, $H \rightarrow \tau\tau$ and $H \rightarrow \gamma\gamma$ channels.}

\FullConference{EPS-HEP 2017, European Physical Society conference on High Energy Physics\\
		5-12 July 2017\\
		Venice, Italy}

\begin{document}

\section{Introduction}

Since the discovery of the Higgs boson by the CMS and ATLAS collaborations \cite{bib:CMS_Higgs, bib:ATLAS_Higgs, bib:CMS_Higgs2}, the measurements of the properties of the particle and its coupling with other particles have come into prominence. The measurement of the top quark Yukawa coupling is of particular interest because of the large mass of the top quark. The production of the Higgs boson is association with a pair of top quarks ($t\bar{t}H$) provides access to the top quark Yukawa coupling directly at tree level. This direct probe, along with indirect probes through studies of Higgs production via gluon fusion, where virtual top quarks provide the dominant standard model (SM) contribution to the loop amplitude can be combined to produce a stringent test of the SM. Moreover, any deviation from SM expectations could be an indication of new physics.

\begin{figure}[htbp]
\centering
\includegraphics[scale=.5]{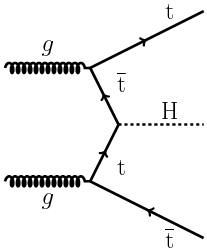}
\caption{
Feynman diagram for $t\bar{t}H$  production at the LHC.
}
\label{fig:ttHFeynman}
\end{figure}

This report presents a summary of the searches for $t\bar{t}H$ using data collected during proton-proton collisions at center of mass energy of 13 TeV in 2016 corresponding to 35.9 $\mathrm{fb^{-1}}$ by the Compact Muon Solenoid (CMS) experiment \cite{bib:CMS} in the different final states corresponding to final states with multileptons (from $H \rightarrow ZZ^{*}$, $H \rightarrow WW^{*}$ and also contribution from $H \rightarrow \tau\tau$ decay channels of the Higgs), final states with hadronically decaying $\tau$ leptons (largely from $H \rightarrow \tau\tau$ decay channels of the Higgs) , and also the $H \rightarrow \gamma\gamma$ final state.

\section{Search for $t\bar{t}H$ in the multilepton final states}

In this channel, the final states of the Higgs boson decay corresponding to $ZZ^{*}$, $WW^{*}$ and also  $\tau\tau$ are targeted. The event selection corresponds to events where at least one of the top quarks decays leptonically, by selecting events with two electrons or muons of the same charge, or with more than three electrons or muons, and hadronic jets compatible with the hadronization of b quarks.

Events are categorised according to the number of leptons, with events with exactly two leptons of the same charge falling into the same-sign dilepton (2LSS) category, events with exactly three leptons falling into the three lepton (3L) category and events with more than three leptons falling into the four lepton category. Events in the 2LSS and 3L channels are further categorized before signal extraction. The 2LSS, category is split according to the events in lepton flavor channels: ee, $\mu\mu$, or e$\mu$. With the exception of the ee channel, events are then further split according to the b-tagging information of the jets in the events. The same categorization for b-tagging criteria is done for 3L events, without splitting in lepton flavors. Finally, to exploit the charge asymmetry in $t\bar{t}W$ production and other background processes, all  2LSS and 3L categories are split according to the sum of lepton charges. Two multivariate discriminants are used in the 2LSS and 3L channels to improve the separation between the signal and  background, particularly the irreducible background of $t\bar{t}W$ and $t\bar{t}Z$ production and from the reducible background of events where at least one of the lepton candidates originates from an hadronic jet (mainly from $t\bar{t}$ +jets), with a complementary matrix element weights based discriminator also used for the 3L channels. In the 4L channel no kinematic discrimination is performed, because of its limited statistical power. The details of the categorisation and the signal extraction procedure are described in Ref. \cite{bib:CMS_PAS_HIG_17_004}. The selection excludes events that contain a hadronically decaying $\tau$ lepton in order to exclude the events entering the analysis with $\tau$ leptons in the final state, which is described in Section~\ref{sec:ttHtau}.

\begin{figure}[htbp]
\centering
\includegraphics[scale=.3]{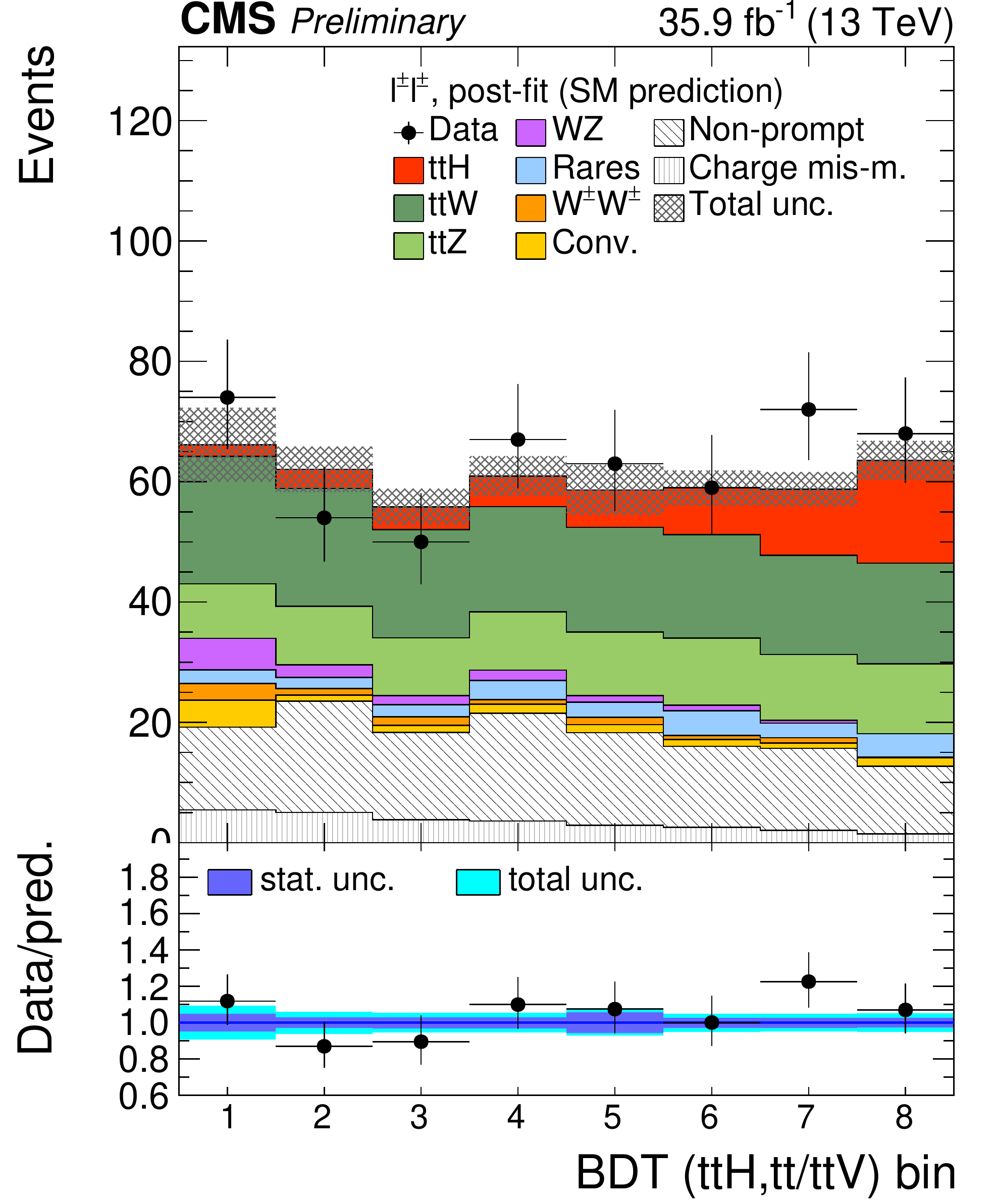}
\includegraphics[scale=.3]{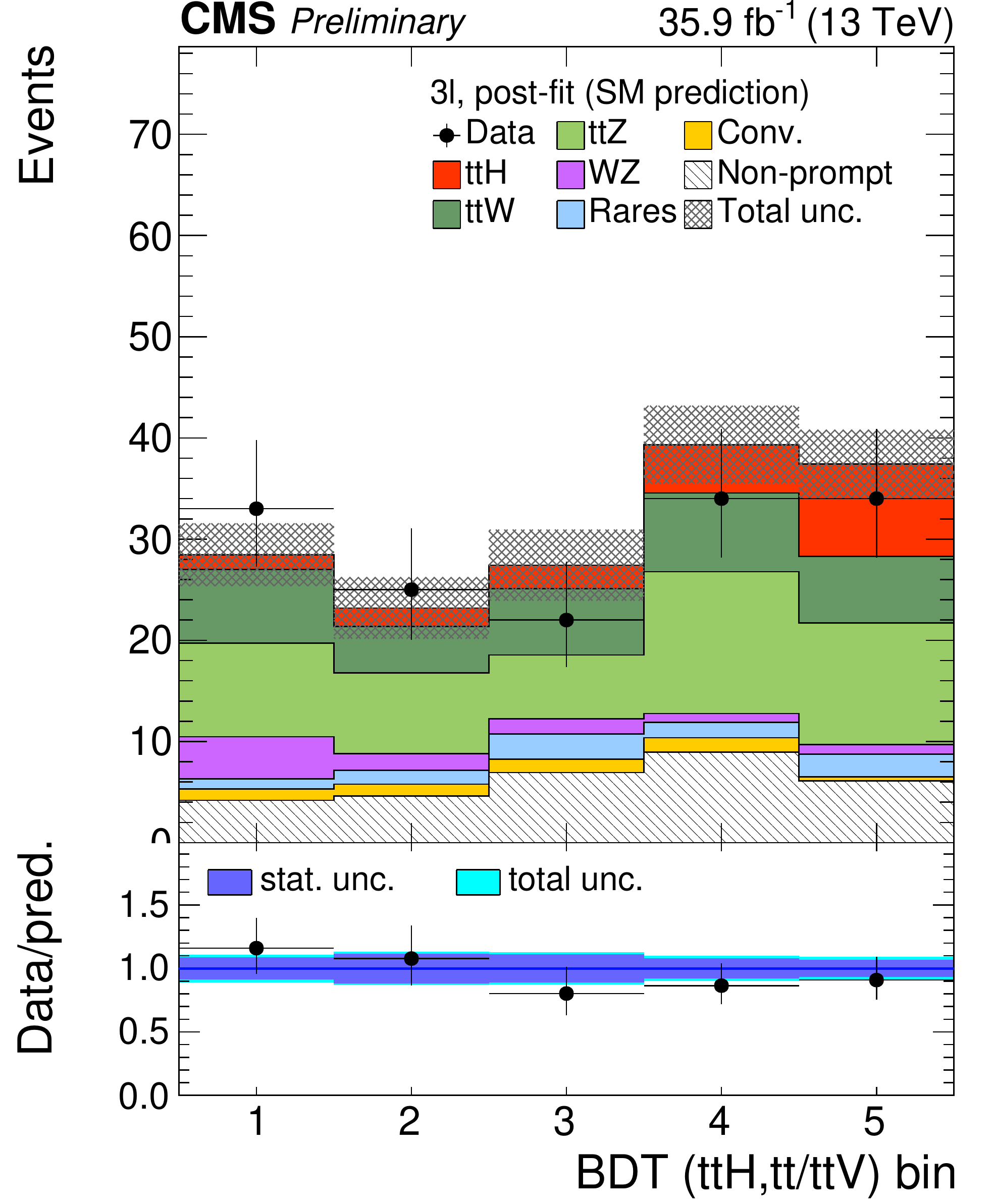}
\includegraphics[scale=.3]{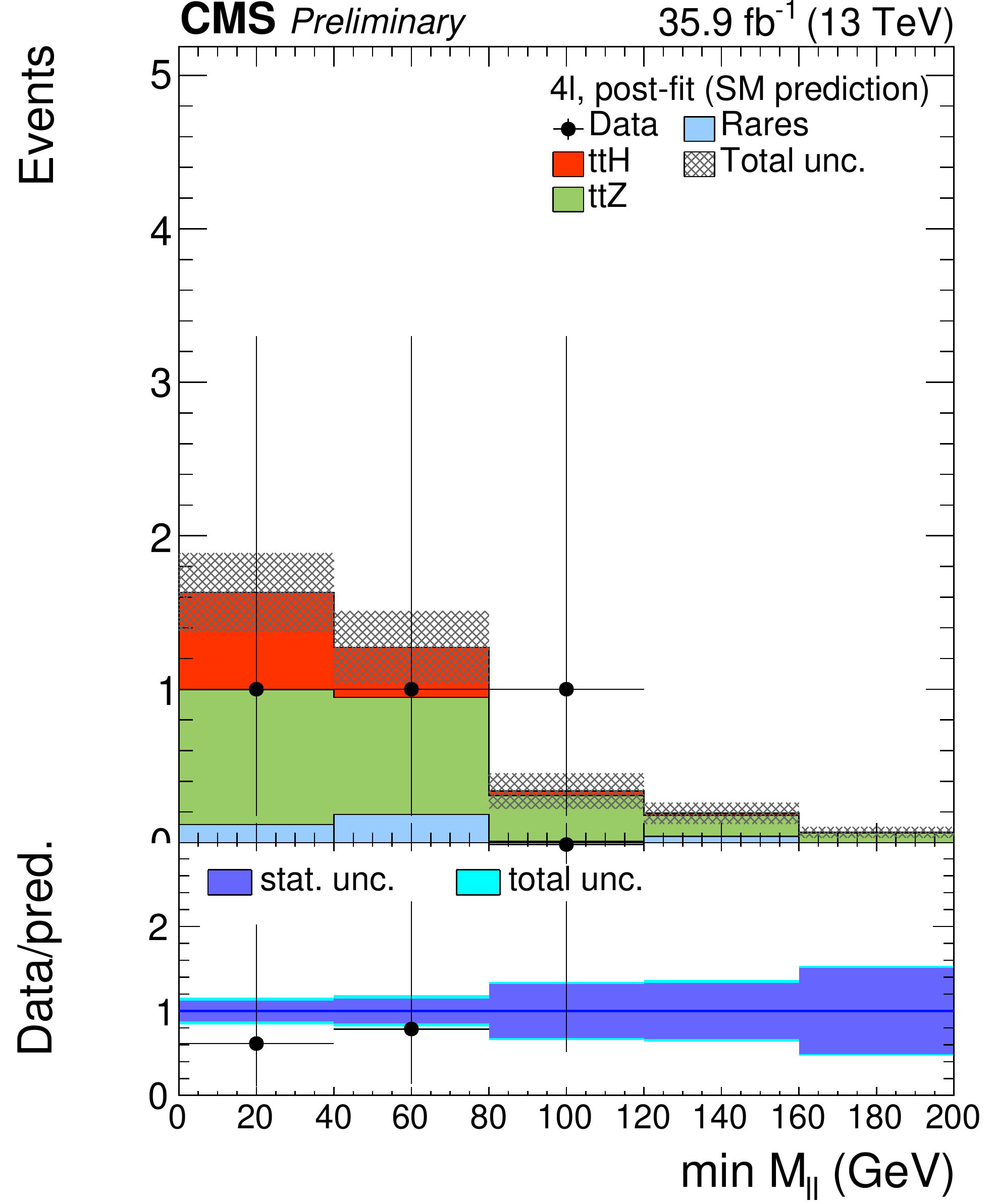}
\caption{
Combination of the BDT classifier outputs in the bins used for signal extraction, for the same-sign dilepton (top left) and three-lepton (top right) channels and the minimum mass of opposite-sign (OS) lepton pairs in the 4L category (bottom). The distributions are shown after the fit to the data, with all processes constrained to the SM expectation. \cite{bib:CMS_PAS_HIG_17_004}
}
\label{fig:ttHmultilepSig}
\end{figure}

The $t\bar{t}H$ yield in the 2LSS and 3L categories is extracted by a simultaneous fit of the two multivariate discriminants. The distribution of the minimum mass of opposite-sign (OS) lepton pairs is used to estimate the signal in the 4L category. The distributions that are used are presented in Figure~\ref{fig:ttHmultilepSig}.

The fitted signal yields are compared with the expectation for a SM Higgs boson of 125 GeV. The signal strength parameter is $\mu = \sigma / \sigma_{SM}$ , acting as a pure scaling of the $t\bar{t}H$ yields.
The observed (expected) best fit signal strength amounts to $1.5^{+0.5}_{-0.5} (1.0^{+0.5}_{-0.4})$, with an observed (expected) significance of 3.3$\sigma$ (2.4$\sigma$) over the background only hypothesis. This result is obtained after combination with the analysis using data collected in 2015 \cite{bib:CMS_PAS_HIG_15_008}. The observed (expected) 95\% CL exclusion limit on $\mu$, in the context of the background only hypothesis, is 2.5 (0.8) \cite{bib:CMS_PAS_HIG_17_004}.

\section{Search for $t\bar{t}H$ in final states with a $\tau$ }\label{sec:ttHtau}

This channel targets $t\bar{t}H$ production with at least one hadronically decaying $\tau$ lepton ($\tau_{H}$) in the final state. This channel is sensitive not only to the $H \rightarrow \tau\tau$ but also to the $H \rightarrow ZZ^{*}$, $H \rightarrow WW^{*}$  decay modes. The event selection is similar to that for the multilepton, with requirements placed on the number of leptons, jets and b-tagged jets, along with the strict requirement of the presence of one or more hadronically decaying $\tau$ leptons.

\begin{figure}[htbp]
\centering
\includegraphics[scale=.3]{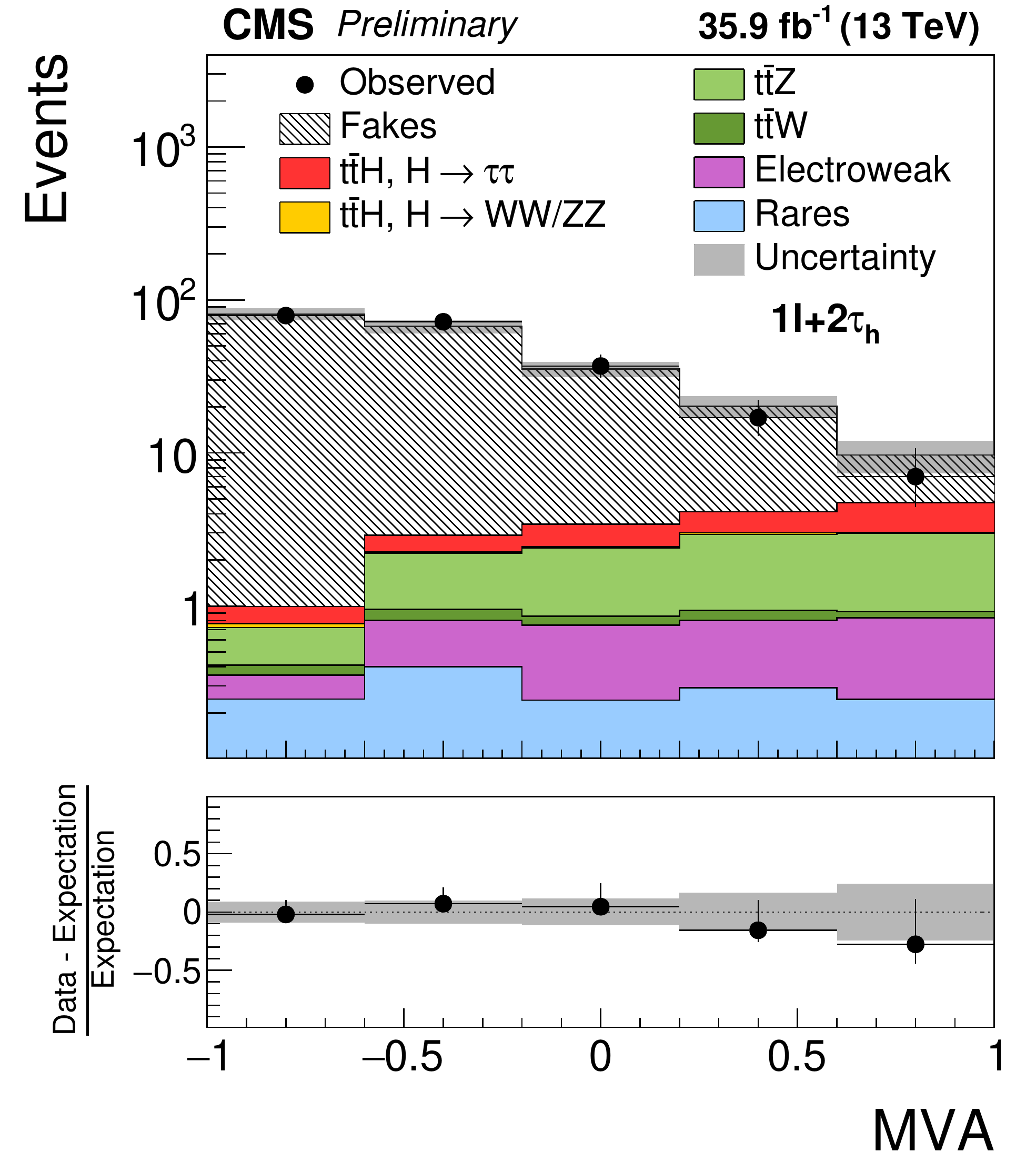}
\includegraphics[scale=.3]{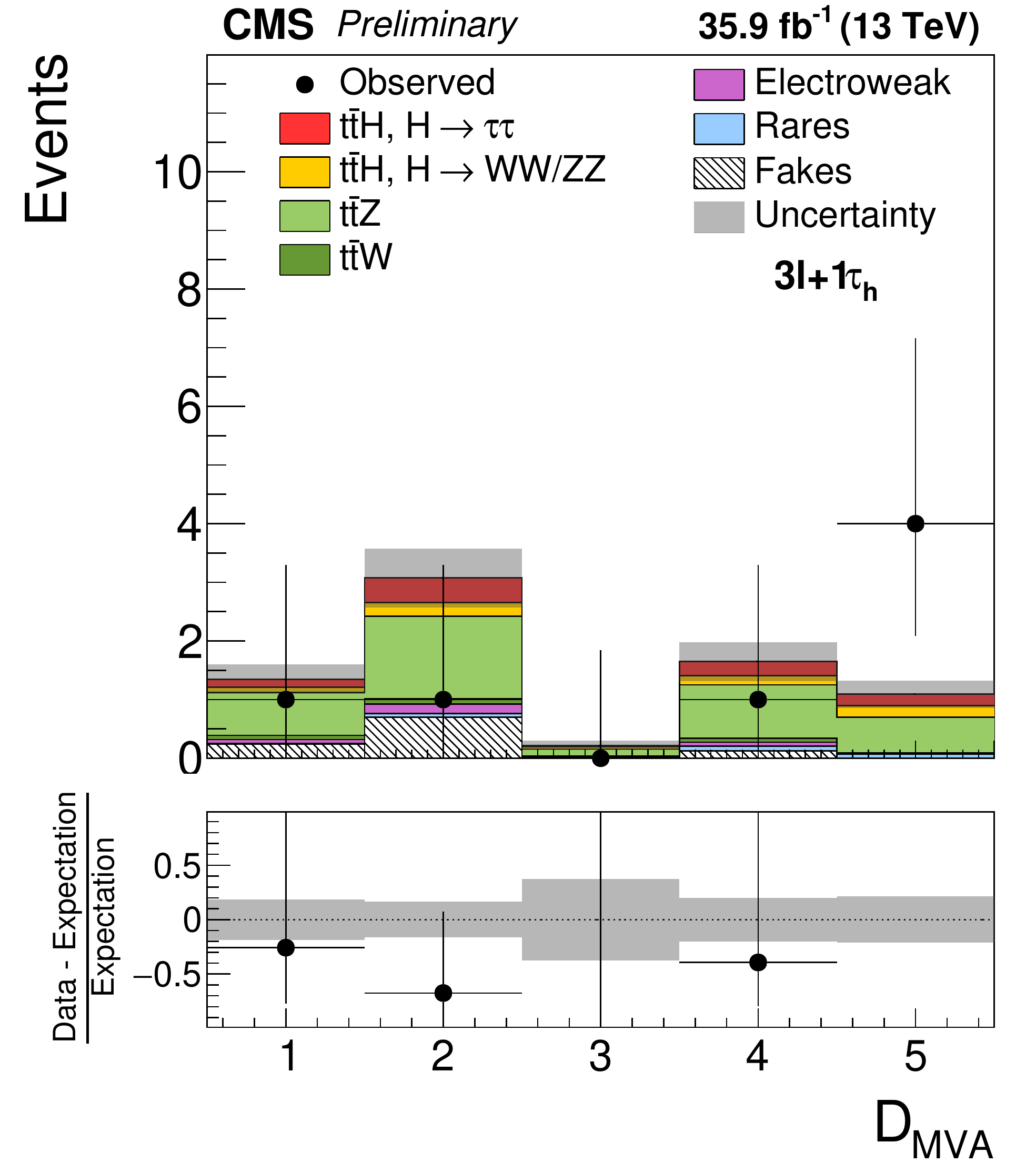}
\\
\includegraphics[scale=.3]{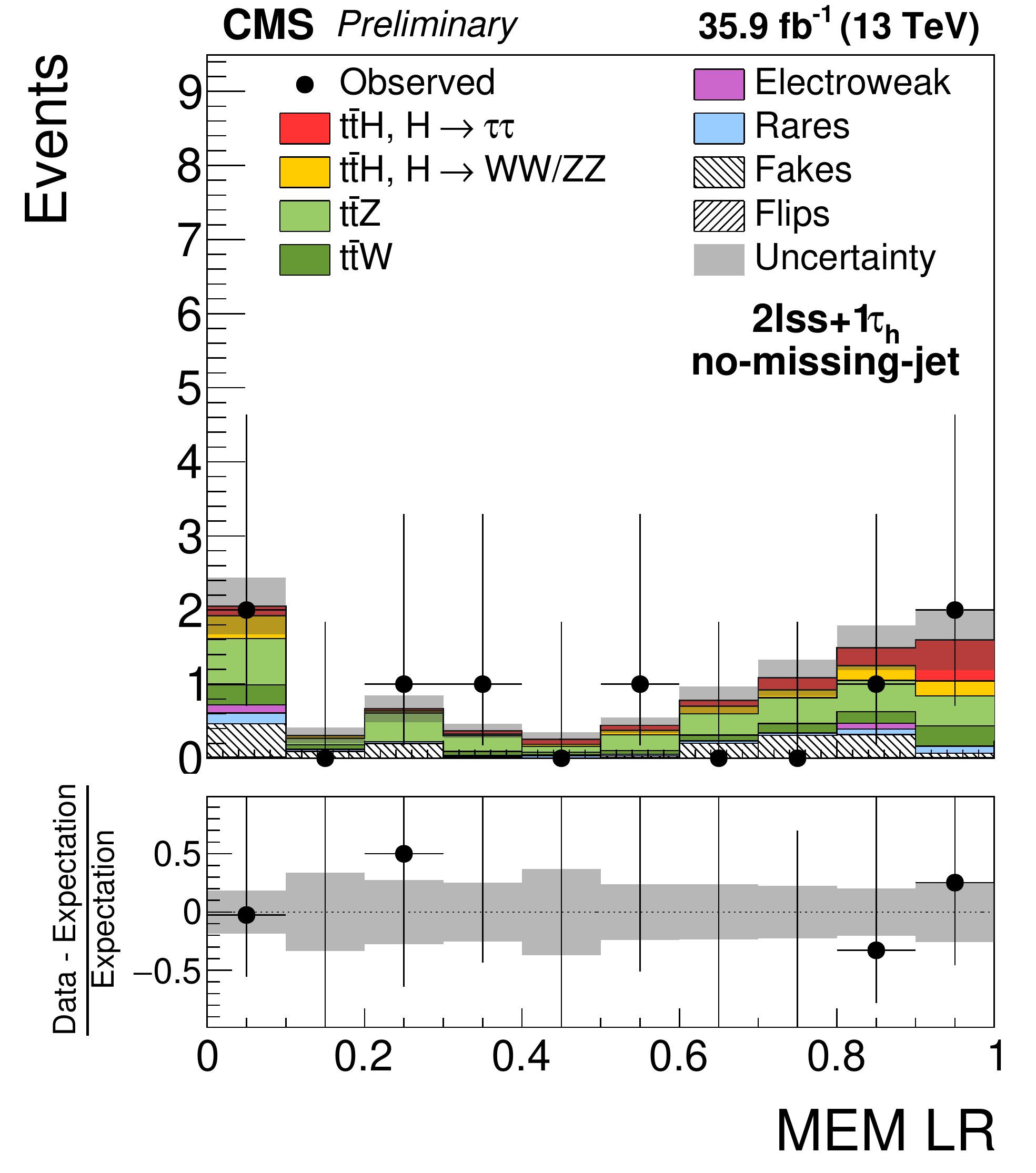}
\includegraphics[scale=.3]{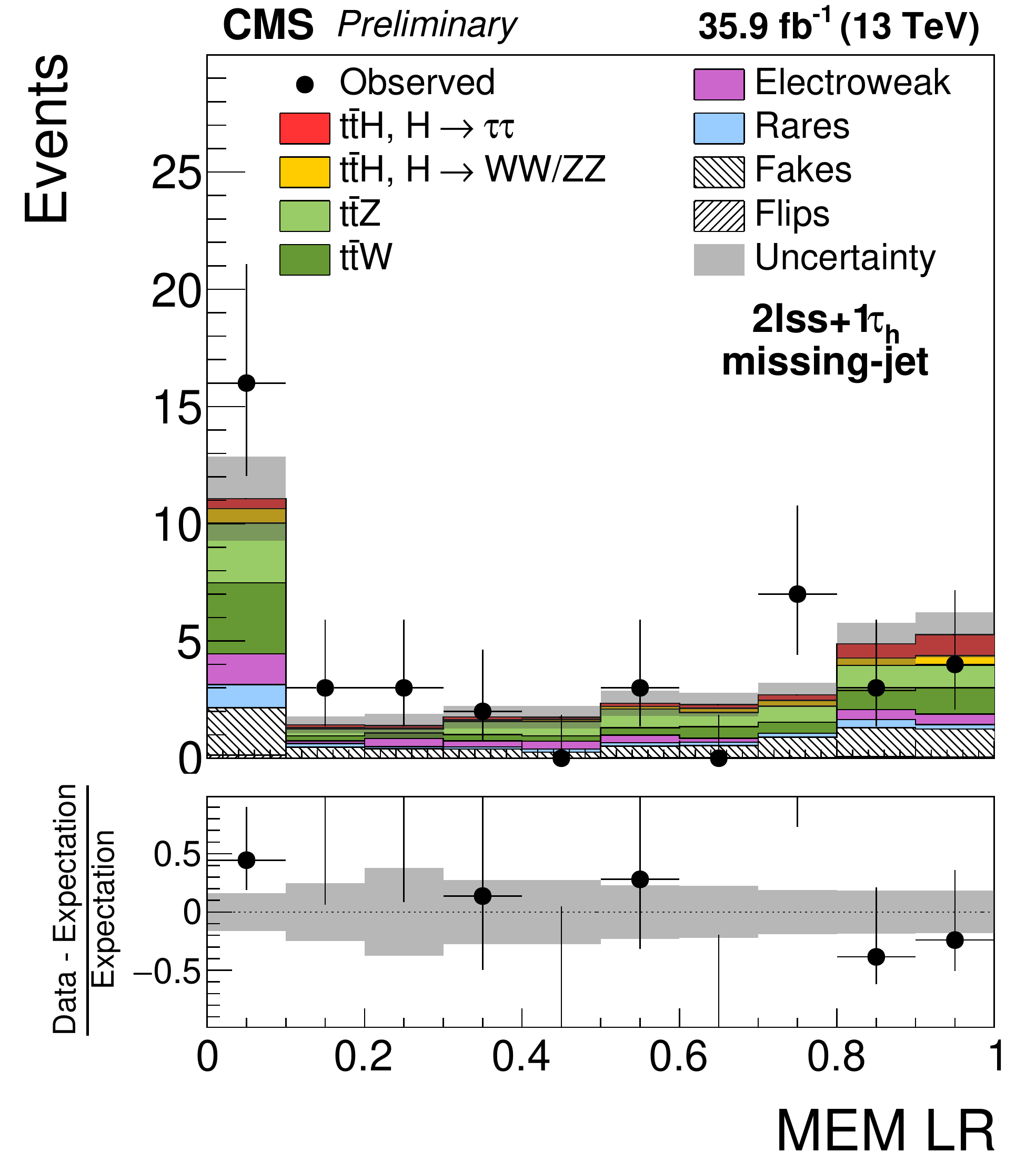}
\caption{
Distributions in the discriminating observables used for the signal extraction in the 1$\ell$ + 2$\tau_{H}$ (top left) and 3$\ell$ + 1$\tau_{H}$ (top right) categories and in the "no-missing-jet" (bottom left) and "missing-jet" (bottom right) subcategories of the 2$\ell$ + 1$\tau_{H}$ category, compared to the SM expectation for the  $t\bar{t}H$ signal and for background processes. The distributions expected for the  $t\bar{t}H$ signal and for the backgrounds are shown for the values of nuisance parameters obtained from the maximum likelihood fit. \cite{bib:CMS_PAS_HIG_17_003}
}
\label{fig:ttHtauSig}
\end{figure}

This analysis is performed in three event categories that are based on the the number of reconstructed leptons and $\tau_{H}$, with the categories being : the one lepton plus two $\tau_{H}$ (1$\ell$ + 2$\tau_{H}$), the two lepton same-sign plus one $\tau_{H}$ (2$\ell$ + 1$\tau_{H}$), and the three lepton plus one $\tau_{H}$ (3$\ell$ + 1$\tau_{H}$) category. The events selected in the 2$\ell$ + 1$\tau_{H}$ category are further divided in two subcategories, the "no-missing-jet" subcategory and the "missing-jet" subcategory, depending on the presence or absence of a pair of jets compatible with the hadronic decay of a W boson coming from the decay of the top quark.
The signal estimation is performed by extracting the signal rate by means of a maximum likelihood fit to the distribution in a discriminating observable in the different categories. In each event category, a different discriminating observable is chosen, in order to achieve the maximal shape separation between the $t\bar{t}H$ signal and background processes. A combination of multivariate techniques including boosted decision trees (BDT) and matrix element method (MEM) analysis are used to create the discriminating observables. Details of the signal extraction procedure are described in Ref. \cite{bib:CMS_PAS_HIG_17_003}. The distributions that are used for signal extraction in the different categories are presented in Figure~\ref{fig:ttHtauSig}.

The results in this channel correspond to a measured signal strength $\mu$ of $0.72^{+0.62}_{-0.53}$, with an observed (expected) significance of 1.4$\sigma$ (1.8$\sigma$) and an upper limit on the signal rate of 2.0 times the SM $t\bar{t}H$ production rate at 95\% CL is set \cite{bib:CMS_PAS_HIG_17_003}.

\section{Search for $t\bar{t}H$ in the $H \rightarrow \gamma\gamma$ final state}

The analysis for $t\bar{t}H$ with $H \rightarrow \gamma\gamma$ is a part of the general $H \rightarrow \gamma\gamma$ analysis as described in Ref. \cite{bib:CMS_PAS_HIG_16_040}. In this channel, a narrow peak corresponding to the the $H \rightarrow \gamma\gamma$ decay is expected on top of a falling diphoton invariant mass ($m_{\gamma\gamma}$) distribution. Diphoton events are selected and the signal estimation is done through by performing fits to the $m_{\gamma\gamma}$ distribution using a signal model derived from simulated signal samples and a data driven estimate of the background. Events are divided into separate categories depending corresponding to signal significance and also based on additional final state objects including leptons, jets and b-tagged jets corresponding to different production modes of the Higgs boson. The signal estimation is performed separately in each category.

\begin{figure}[htbp]
\centering
\includegraphics[scale=.35]{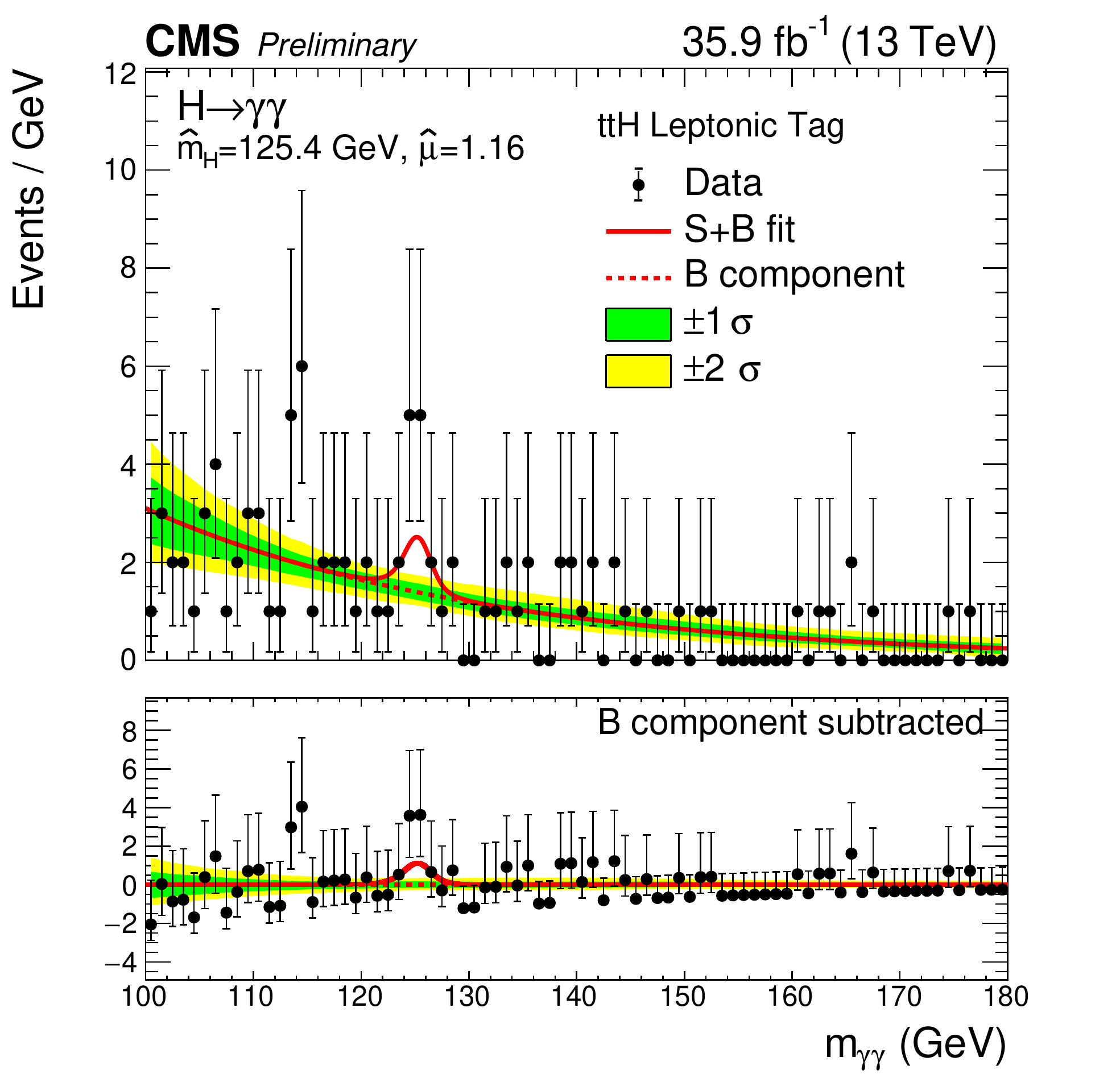}
\includegraphics[scale=.35]{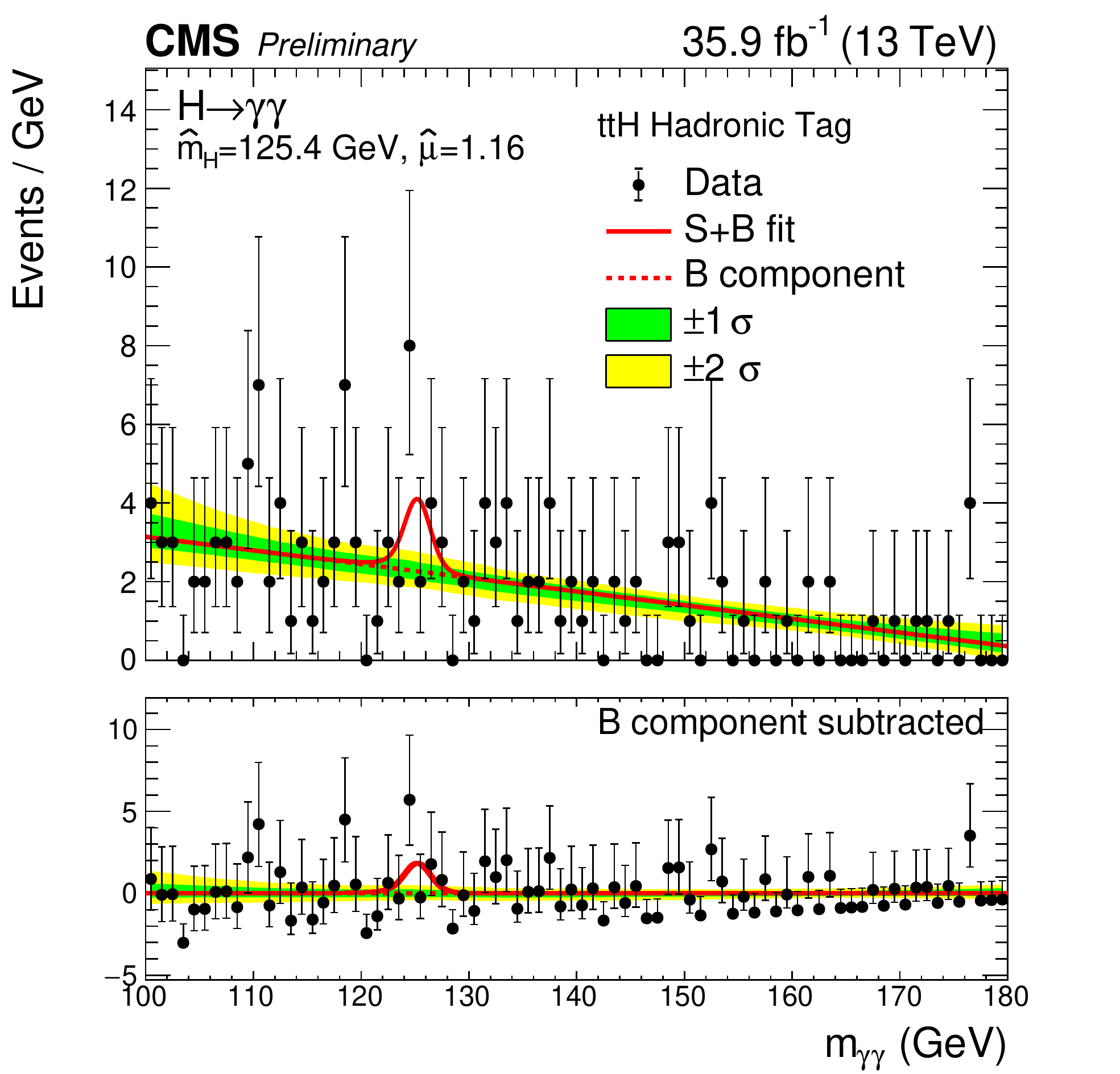}
\caption{
The $m_{\gamma\gamma}$ distribution for the $t\bar{t}H$ Leptonic category (left) and the $t\bar{t}H$ Hadronic category (right). The one standard deviation (green) and two standard deviation bands (yellow) include the uncertainties in the background component of the fit. The bottom plot shows the residuals after background subtraction. \cite{bib:CMS_PAS_HIG_16_040}
}
\label{fig:ttHGam}
\end{figure}

Two categories correspond to the $t\bar{t}H$ production based, one corresponding to the semi-leptonic decay of the top quarks ($t\bar{t}H$ Leptonic) and the other corresponding to the hadronic decay of the top quarks ($t\bar{t}H$ Hadronic) \cite{bib:CMS_PAS_HIG_16_040}. The fits to the $m_{\gamma\gamma}$ distributions in the $t\bar{t}H$ tagged categories are shown in Figure~\ref{fig:ttHGam}.

The results in the $H \rightarrow \gamma\gamma$ channel for the $t\bar{t}H$ production mode correspond to a measured signal strength $\mu$ of $2.2^{+0.9}_{-0.8}$, with an observed (expected) significance of 3.3$\sigma$ (1.5$\sigma$) \cite{bib:CMS_PAS_HIG_16_040}.

\section{Summary}

The searches for the production of the Higgs boson in association with a pair of top quarks using data collected during proton-proton collisions at center of mass energy of 13 TeV in 2016 corresponding to 35.9 $\mathrm{fb^{-1}}$ by the CMS experiment are presented for the multilepton channel, $\tau_{H}$ channel and the diphoton channel. The measurements are compatible with the SM predictions, with the observed (expected) significance measured to be 3.3$\sigma$ (2.5 $\sigma$) for the analysis with the multilepton final states \cite{bib:CMS_PAS_HIG_17_004}, 1.4$\sigma$ (1.8 $\sigma$)  for the analysis with the $\tau_{H}$ final state \cite{bib:CMS_PAS_HIG_17_003}, and 3.3$\sigma$ (1.5 $\sigma$)  for the analysis with the $\gamma\gamma$ final state \cite{bib:CMS_PAS_HIG_16_040}.

\end{document}